\documentclass[12pt]{article}

\usepackage{ a4, graphicx, amsmath, amsthm, color, ulem, psfrag, url, eurosym}

\newcommand { \be }{ \begin{equation} }
\newcommand {\ee} {\end{equation}}
\newcommand {\bmath} {\begin {displaymath} }
\newcommand {\emath} {\end {displaymath} }

\newtheoremstyle{mytheoremstyle}{3pt}{3pt}{\itshape}{}{\scshape}{:}{0.5em}{}
\theoremstyle{mytheoremstyle}

\usepackage[comma,longnamesfirst]{natbib}
\begin{document}

\normalem

\title{\vspace*{-0.7in}
\textbf{A Bayesian Hierarchical Model for Comparative Evaluation of Teaching Quality Indicators in Higher Education}}

\author{
D.~Fouskakis\thanks{D.~Fouskakis is with the Department of Mathematics,
National Technical University of Athens, Zografou Campus, Athens 15780
Greece; email \texttt{fouskakis@math.ntua.gr}}, \ G. Petrakos\thanks{G.~Petrakos is with the Department of Public Administration,
Panteion University of Social and Political Sciences, 136 Syngrou Avenue, 17671, Athens, Greece; email \texttt{petrakos@panteion.gr}} \ and 
I. Vavouras\thanks{I. Vavouras is with the Department of Public Administration,
Panteion University of Social and Political Sciences, 136 Syngrou Avenue, 17671, Athens, Greece; email \texttt{vavouras@panteion.gr}}
}

\date{}

\maketitle

\vspace*{-0.3in}

\noindent
\textbf{Summary:} The problem motivating the paper is the quantification of students’ preferences regarding teaching/coursework quality, under certain numerical restrictions, in order to build a model for identifying, assessing and monitoring the major components of the overall academic quality. After reviewing the strengths and limitations of conjoint analysis and of the random coefficient regression model used in similar problems in the past, we propose a Bayesian beta regression model with a Dirichlet prior on the model coefficients. This approach not only allows for the incorporation of informative prior when it is available but also provides user friendly interfaces and direct probability interpretations for all quantities. Furthermore, it is a natural way to implement the usual constraints for the model weights/coefficients. This model was applied to data collected in 2009 and 2013 from undergraduate students in Panteion University, Athens, Greece and besides the construction of an instrument for the assessment and monitoring of teaching quality, it gave some input for a preliminary discussion on the association of the differences in students preferences between the two time periods with the current Greek economic and financial crisis.

\vspace*{0.15in}

\section{Introduction}

The institutions of higher education worldwide are generally independent organizations. In Greece they are self-administered. In all cases external assessment and self-assessment are critical processes for maintaining, securing and improving the quality of the services offered. The evaluation is usually based upon four large groups of criteria, which comprise the major pillars of academic quality. They refer to teaching, research, academic program and supportive organizational facilities and services. 

The core issue of the academic evaluation process is the assessment of the quality of an individual course/instructor by the students attending this course. By filling in a relevant and well prepared questionnaire, students evaluate, in a measurement scale, certain quality attributes, concerning the course (content, text book, grading system) and the instructor (teaching, knowledge, behaviour) that are widely considered as being of major importance. Several instruments for evaluating teaching performance have been developed and used internationally, just mentioning Feldman's categories \citep{f1976} and SEEQ's (Students Evaluation of Educational Quality) factors \citep{m1982}. 

The aggregation of the results in order to reach some reliable conclusions referring to the objectives of the evaluation, inevitably involves the assignment of weights to the attributes/variables at a certain point of the statistical analysis. This is usually done by subject matter experts, that subjectively (most of the times uniformly) assign weights to those variables. For an extensive review of different weighting schemes in student ratings see \cite{hdb2004} and \cite{m2007}. It is worth mentioning here that, by assigning weights uniformly, the analysis follows the weighting scheme of the questionnaire design, also made by (other) experts. This commonly used practice rises a major criticism since the participants (population representatives) do not have the opportunity to contribute to the weighting assignment. To fully understand the consequences of this methodological weakness, one may consider that: 
\begin{itemize}
\item [a)]	The weighting process is critical for the results of the statistical analysis and therefore for the outcome of the whole project and its policy implications. It should not be left therefore entirely to those conducting the evaluation, since in this case it is not guaranteed that the outcomes of the evaluation are objective and not manipulated by those carrying out the assessment. By changing the relative weights one can increase or reduce the significance of certain attributes affecting critically the outcomes of the evaluation process. In that case the evaluation process itself becomes unreliable and can be very easily rejected.
\item[b)]	Students, the statistical survey units, comprise the population to which the assessment is basically addressed, and moreover the social group that will be mostly affected by the results of this assessment, being at the receiving end of the academic services offered. This final end of the assessment, that is the welfare improvement of this group, can not be ignored at any stage of the evaluation.
\end{itemize}

If these methodological weaknesses prevail in our analysis, there is no guarantee that the evaluation really contributes to the improvement of the efficiency of the academic units relative to the expectations or demands of the social groups to which it is addressed. This issue arises, since the analysis is possible to give emphasis to the advancement of targets (characteristics) that the affected social groups do not consider as being of major importance and on the contrary not to give emphasis (practically to ignore) to the advancement of targets (characteristics) that the affected social groups consider as being of critical importance.

More generally, if we accept that the individual is the best judge of his/her own welfare and not someone else, and to the extent that the notion of methodological individualism is also accepted and consequently the social welfare is the outcome of the integration of the individual welfare functions (as functions of the individual characteristics) and nothing else, the evaluation methodology that ignores the issues we have raised above is not compatible with the conditions of the best social choice rule. We must note that according to methodological individualism the social groups of any form are nothing more than sums  of individuals and as a result the characteristics of these social groups can only be analysed on the basis of the individual characteristics. For an analysis of the issue, see mainly \cite{a1994}, \cite{h1948}, \cite{h1998}, \cite{h2007} and \cite{w1952}.

Understanding therefore how university students perceive and evaluate the various components of the coursework and teaching offered by their lecturers and as a result by their institution is critical for instructors and decision makers. The above reasoning can serve as a baseline for the continuous endeavour to improve, design and redesign a competitive academic program.

An appropriate approach to elicit population preferences is conjoint analysis (CA) that has been used for more than 40 years to model and measure consumer's perception of product or service quality. It appeared as conjoint measurement in psychology \citep{lt1964} and it was successfully transferred to marketing applications in early seventies by \cite{gr1971} and \cite{j1974}. The basic concept behind CA is that in a well-designed experiment, respondents, that actually constitute a random sample of the possible customer population, are asked to evaluate products or services as combinations of their main attributes/characteristics. These individual pieces of information are combined through the statistical analysis and result in a quantitative estimation of customer preferences usually as a vector of attribute weights. According to \cite{hr2004}, ``the essence of CA is to identify and measure a mapping from more detailed descriptions of a product or service onto an overall measure of the customer's  evaluation of that product". This mapping, realized upon various experimental designs, using different multivariate statistical techniques, is the primary objective of almost all theoretical and applied work in CA (\cite{claabhjksstw1994}; \cite{gs1990}; \cite{o2010}).  

From the early days of CA, various types of stochastic models such as ordinary least squares regression models, as well as logit, probit, multinomial probit and nested logit or probit  models used with linear, quadratic or log response function \citep{m1974} were applied to conjoint data. In the nineties, Hierarchical Bayes (HB) models, became very popular in CA since they improve the reliability and predictive validity of CA and provide robust estimates \citep{o2010}. In principal, HB is a two level approach with the consideration of a prior multivariate distribution for individual parameters (part worths) at the higher level and a generalised linear model delivering the probabilities of possible outcomes (preferences) given the above mention priors, at the lower level. HB borrows information from other respondents to stabilize maximum likelihood parameter estimation for each individual, using iterative procedures. HB approach provides stable and accurate models that can recover heterogeneity by separating it from noise even when observations are less than estimated parameters \citep{lwgy1996}.

When personal computers became powerful and fast in data processing several optimisation methods were developed and used in CA. The main families of these optimisation methods are Polyhedron Center Estimation (\cite{tshd2003}; \cite{ths2004}), Support Vector Machines \citep{ebz2002} and finally Genetic Algorithms \citep{h1975}, where attributes considered as genes, so the next generation of products or services will be identified, evaluated and finally introduced. 

While main characteristics of product quality, such as durability and number of defeats are objectively measured \citep{g1983}, main common features in service quality such as intangibility, heterogeneity and inseparability of production and consumptions cannot \citep{pzb1985}.   Service quality is often defined as the discrepancy between consumers' perception of services offered by a certain organization and their expectations about the organization offering those services \citep{pzb1988}.  Conceptualizing service quality, researchers define pairs of distinct terms like perceived vs. objective quality, satisfaction vs. quality and perceptions vs. expectations. \cite{hc1985} came to the conclusion that customers do not perceive quality the same way researchers do. In extensive focus group interviews \citep{pzb1985} came up with several instances, where customers were satisfied with a certain service but the corresponding institution or firm quality was not highly ranked. There have been numerous applications of CA in various service industries such as accommodation and haircut services \citep{oi1995}, transportation \citep{h2001}, telecommunication (\cite{k2004}; \cite{kccp2008}), health (\cite{kvm2012}; \cite{rf2000}) and banking \citep{mhd2013}.

Focusing on education, \cite{z1983} launched a conjoint experiment in order to optimize the design of a new course in quantitative marketing addressed to MBA students at the University of South California. In another conjoint experiment in the University of South Dakota, \cite{pccfflmpprssw2007} applied a 24 full factorial design to both graduate and undergraduate students reporting differences in preference between the two groups. Furthermore, differences in students' preference regarding teaching attributes were found between those who were more and those who were less satisfied with the academic program. \cite{zmm2006} propose a four attribute CA in a $4 \times 3 \times 2 \times 2$ design reduced to $1/3$ using SPSS-CA procedure in order to design courses in environmental entrepreneurship education programs. In \cite{wb2009} six course attributes in sport management undergraduate students were evaluated using classical CA and OLS, followed by sensitivity analysis. 

Recently, there is an increasing volume of studies using CA for quality assessment in education \citep{kspm2012} to evaluate university teaching considering heterogeneity of students' preferences. \cite{kss2009} used CA to determine the most influential attributes of English Medium Instruction (EMI) classes. Furthermore, \cite{aaa12} and \cite{l2013} used SERVQUAL methodology \citep{pzb1988} and CA to assess quality in educational institutions in Iran and Italy, respectively. A serious limitation in all conjoint applications to teaching/course evaluation is the large number of attributes that comprise the academic services that cannot be drastically reduced. Most researchers test only some attributes that do not add up to complete teaching scenarios and/or use fractional factorial designs with limitations in estimation capabilities.  

An alternative approach was introduced by \cite{py2006} aiming at the construction of a model for building a linear quality index for service evaluation. According to this methodology each statistical unit was asked to evaluate both the quality attributes/characteristics and the overall (global) service quality. Then a Dirichlet random coefficient regression model was used, in order to obtain the weights of the attributes and finally to construct the quality index. This model includes a) the restriction that the weights must be positive and add up to one and b) a latent variable that incorporates all quality attributes not considered in the survey. The design and the model of this paper were used to measure the quality of the rail road system in Spain.

Using a similar experimental design, we have asked university students to evaluate both the quality of certain academic attributes concerning courses teaching and content and the overall courses quality. By gathering this type of data, we aim to cream off the variables' weights from the students' mind, since every survey participant evaluates some attributes as more important than others by giving them scores similar to one given to the overall quality. Then, we fit a Bayesian beta regression model with a uniform Dirichlet prior on weights in order to derive the posterior distribution of attribute weights, subject to the usual constraints, and construct a global quality index for teaching based on students' preferences. This model can be used as an instrument for the evaluation of individual course/instructor quality incorporating student data gathered through the regular process for academic evaluation of each course in the program.
The outcome of the analysis can be also interpreted and used as a general assessment of the global and the partial quality of academic program/teaching in the specific academic unit at a certain period of time. Since the study was repeated at two different time periods (2009 and 2013), changes can be tracked, analysed and associated with changes in the academic and/or general socio economic environment. Taking into consideration the major changes in the socio economic environment in Greece due to the present financial and economic crisis, we will discuss possible associations with the results of our study. 

\section{Survey Design and Metadata}

The survey was addressed to undergraduate students with at least two years of academic experience, majoring Public Administration at Panteion University of Social and Political Science, in Athens, Greece. Therefore the population under study comprise of university students having attended more than twenty courses, subjecting various academic disciplines, taught by different instructors. 

Initially, a random sample of 96 students participated in the survey was drawn in May 2009, while a second sample of 95 students, responding to the same questionnaire, was obtained four years later, in May 2013. Besides the survey instrument, the experimental conditions as well as the oral and written guidelines were also identical for both survey instances. It should be pointed out that since 2008, in a time frame within both survey instances occurred, no substantial changes have been made either in the academic program or in the teaching stuff of the department.  Considering the survey quality, unit non response rate was below 10\% and item non response was negligible in both instances. 
 
Students were asked to evaluate certain quality attributes regarding all courses taken so far in their academic program. More specifically, they were asked to fill in the percentage of courses/ instructors having positive quality characteristics, based on their academic view and experience (i.e. in what percentage of your courses, taken so far, the instructor has developed good communication channels with the students, encouraging questions and dialogue?). The scale of measurement was ordinal with 11 levels corresponding to sequential, 10\% width percentage intervals, from 0 to 100.

The quality of the (course/instructor) academic module was measured by a set of twenty attributes structured in a three level hierarchical model. At level I, we identify two large groups of quality attributes, one referring to the courses and another to instructors. Within these groups, at level II, general attributes, such as content of the course, text book and grading procedure for the first group, as well as, teaching, knowledge and behaviour of the instructor for the second group, are included as main research questions \citep{bl2003}. Finally, at level III, research questions break down to individual quality attributes, translated into twenty single item questions in the survey instrument. The questionnaire was completed with a last question asking for an overall quality evaluation of the general course/teaching module, in a comparable measurement scale. 

The structure, content and wording of the questionnaire were based on prototypes and guidelines from \cite{e2009} and \cite{he2007}, Feldman's categories \citep{f1976} and SEEQ's factors \citep{m1982}. The twenty individual quality components are presented, in the same order as the corresponding questions appeared in the questionnaire, in Table \ref{qc}.

\begin{table}[ht!]
\small
\caption{List of teaching/coursework quality components}
\label{qc}
\begin{center}
\begin{tabular}{ll}
\hline Q(I)	& Description \\
\hline                                                 
 1&	Clarity of the course objectives ($w_{1}$)\\
2&	Relevance of course content to course objectives ($w_{2}$)\\
3&	Relevance of course content to established standards ($w_{3}$)\\
4&	Value of course materials ($w_{4}$)\\
5&	Course content met academic and professional requirements ($w_{5}$)\\
6&	Importance of course attendance towards mastering and successfully completing the course ($w_{6}$)\\
7&	Examination's subject and means compatibility with course content and teaching ($w_{7}$)\\
8&	Examination's difficulty compatible with course difficulty ($w_{8}$)\\
9&	Examination's means used contribution to student skills development ($w_{9}$)\\
10&	Fairness, impartiality of examination ($w_{10}$)\\
11&	Instructor's course organization ($w_{11}$)\\
12&	Instructor's preparation ($w_{12}$)\\
13&	Attractiveness of lecturing ($w_{13}$)\\
14&	Instructor's communicability ($w_{14}$)\\
15&	Instructor's subject knowledge ($w_{15}$)\\
16&	Communication  with students, encouragement of group interaction ($w_{16}$)\\ 
17&	Consistency in teaching program and availability in office hours ($w_{17}$)\\
18&	Fair treatment of students ($w_{18}$)\\
19&	Respect for students ($w_{19}$)\\
20&	Intellectual challenge and expansiveness to current scientific and actual trends ($w_{20}$)\\
21&	Overall teaching/coursework evaluation ($y$)\\      
\hline
\end{tabular}
\end{center}
\end{table}

\normalsize

A pre-testing version of the questionnaire was released, answered and evaluated by a group of 6 volunteers, coming from the population under study, resulted in several changes in wording and ordering of the questions in the final questionnaire.

\section{Bayesian Hierarchical Model}

Suppose that a random sample of $n_j$ students has been surveyed, with $j=1,2$ denoting the two different time periods. Let $n=n_1+n_2$ be the total sample size and $y_{ij}$ denoting the observed value of the response variable (overall teaching/coursework evaluation score) given by the $i^{th}$ individual on time $j$ ($i=1,\dots, n_j$). Furthermore let $\mathbf{x}_{ij}=(x_{ij,1},\dots,x_{ij,k})^T$ denoting the observed value for the vector of the $k$ known attributes. Additionally, let $z_{ij}$ denoting the unobserved random variable (latent variable) corresponding to the evaluation of the unspecified factors for student $i$ at time $j$. Without loss of generality we assume that the data has been scaled and therefore all the above mentioned variables are scores that take values in $[0,1]$. 

We assume that the response variable follows a beta distribution with mean that is modelled as a weighted linear combination of the observed and unobserved attributes and constant variance. Therefore the coefficients of the model can be interpreted as weights (i.e. positive random variables that add to one) and thus they measure the relative importance that student $i$ at time $j$ gives to the different attributes and the latent variable.   

The observed data for the response variable were initially percentages, indicating the overall evaluation score.  
In the literature, various methods to model
percentage data have been proposed. A well-known strategy is
to transform the percentage outcome and to carry out ordinary least square 
regression. The arcsine square root and the logit functions are among the most typical examples of such transformations. 
These models are quite popular among analysts but quite often their assumptions are not met; see for example \cite{warton-hui-2011}. 
An alternative approach is to use regression models that are based on the binomial
distribution. Unfortunately, quite often, percentage outcomes that are based on the binomial model
are overdispersed, meaning that they show a larger
variability than expected by the binomial distribution. In addition, there are numerous applications
where percentage outcomes are non-binomial (e.g. fractions of communities with unknown size). To overcome the aforementioned problems and limitations, a beta regression model is often used \citep{ferrari-cribari-neto-2004}. Because the beta distribution has a highly flexible
shape, it is suitable to represent arbitrary outcome variables
measured on the percentage scale. Additionally, even simple beta regression models conveniently account for overdispersion by
including a precision parameter in order to adjust the conditional variance of the percentage outcome. 
In a similar problem like the one here, \cite{carmichael_2006} compared the fit of several models, including the beta regression, to model
educational performance data and concluded that a linear model based upon the beta distribution is usually the most appropriate.

A frequentist approach to beta regression was first developed by \cite{ferrari-cribari-neto-2004}, where asymptotic properties of maximum likelihood estimators were used to obtain inferences. In contrast, inferences based on a Bayesian analysis do not rely
on large-sample approximations. Moreover, a Bayesian approach
not only allows for the incorporation of informative prior information when it is available,
but also provides user-friendly inferences and direct probability interpretations for all
quantities. For an example of a Bayesian beta regression model see \cite{branscum_johnson_thurmond_2007}.

The proposed model, for the two different time periods ($j=1,2$), is a beta regression and has the following form:

\begin{eqnarray} \label{sep}
Y_{ij} &\stackrel{ \mbox{\footnotesize indep } }{ \sim }& Beta(a_{ij}, b_{ij}) \nonumber \\ 
a_{ij} &=& \frac{(1-\mu_{ij})\mu_{ij}^2- \mu_{ij} \sigma^2_{j}}{\sigma^2_{j}} \nonumber \\
b_{ij} &=& \frac{(1-\mu_{ij})(\mu_{ij}-\mu_{ij}^2-\sigma^2_{j})}{\sigma^2_{j}} \nonumber \\
\mu_{ij} &=& w_{j,1}x_{ij,1}+\dots + w_{j,k}x_{ij,k}+ w_{j,k+1}z_{ij}
\end{eqnarray}
where $\mu_{ij}$ denotes the mean of the response variable $Y_{ij}$ (overall evaluation score) and $\sigma^2_{j}$ the variance of the response variable. Furthermore $\mathbf{w}_{j}=(w_{j,1},\dots,w_{j,k},w_{j,k+1})^T$ is the random vector of weights measuring the relative importance at time $j$ of the 
different attributes and the latent variable. We have that $0 \leq w_{j,\ell} \leq 1$ ($\ell=1,\dots,k+1$, $j=1,2$) and $\sum_{\ell=1}^{k+1} w_{j,\ell}=1$ for $j=1,2$. Note that the linear prediction in model (\ref{sep}) take values in $[0,1]$, thus the identical link function is used for interpretation reasons. Additionally, to keep the model simple, we have assumed a constant variance. With this choice, we still account for overdispresion, but due to the absent of additional covariates in the model (e.g. age or gender of students) we assume that this overdispresion does not depend on the values of the predictors. 

We use a uniform Dirichlet prior on the weights, i.e. $\mathbf{w}_{j} \sim Dirichlet(1,\dots,1)$ and a uniform prior on $\sigma^2_{j}$: $\sigma^2_{j} \sim Unif(0,m)$, with $m=min \{(1-\mu_{ij})\mu_{ij}\}$. The first prior is the usual ``low information'' prior on weights, while with the second prior we make sure that $a_{ij}, b_{ij} \geq 0$. For the latent variables we assume that they are independent and they come from a beta distribution with both parameters equal to 0.5. 

We fit model (\ref{sep}) initially (for $j=1,2$) and we compare its predictive ability with the one from the  
simpler (joint) model ($i=1,\dots,n$):
\begin{eqnarray} \label{joint}
Y_{i} &\stackrel{ \mbox{\footnotesize indep } }{ \sim }& Beta(a_{i}, b_{i}) \nonumber \\ 
a_{i} &=& \frac{(1-\mu_{i})\mu_{i}^2- \mu_{i} \sigma^2}{\sigma^2} \nonumber \\
b_{i} &=& \frac{(1-\mu_{i})(\mu_{i}-\mu_{i}^2-\sigma^2)}{\sigma^2} \nonumber \\
\mu_{i} &=& w_{1}x_{i,1}+\dots + w_{k}x_{i,k}+ w_{k+1}z_{i}
\end{eqnarray}
where now $\mu_{i}$ denotes the mean of the random variable $Y_{i}$ and $\sigma^2$ the (common) variance of $Y_{i}$. Furthermore $\mathbf{w}=(w_{1},\dots,w_{k},w_{k+1})^T$ is the random vector of weights measuring the relative importance of the  
different attributes and the latent variable. Once again we have that $0 \leq w_{\ell} \leq 1$ ($\ell=1,\dots,k+1$) with $\sum_{\ell=1}^{k+1} w_{\ell}=1$ and the identical link function is used. 
As before, we use a uniform Dirichlet prior on the weights, i.e. $\mathbf{w} \sim Dirichlet(1,\dots,1)$ and a-priori we assume that $\sigma^2 \sim Unif(0,m)$, with $m=min \{(1-\mu_{i})\mu_{i}\}$. Finally we assume again that the latent variables are independent and come from a beta distribution with both parameters 0.5.

We use WinBugs to fit the above mentioned models. Note that the above models assumes that responses are strictly larger than 0 and
strictly smaller than 1. In our dataset we removed two observations that did not meet this requirement.

\section {Results}

\begin{figure}[ht!]
\caption{Boxplots summarizing posterior distributions of the model weights (1: First Period , 2: Second Period, 3: Joint Data)}
\label{comp}
\vspace{-2em}
\begin{center}
\psfrag{w1}[c][c][0.9]{$w_{1}$}
\psfrag{w2}[c][c][0.9]{$w_{2}$}
\psfrag{w3}[c][c][0.9]{$w_{3}$}
\psfrag{w4}[c][c][0.9]{$w_{4}$}
\psfrag{w5}[c][c][0.9]{$w_{5}$}
\psfrag{w6}[c][c][0.9]{$w_{6}$}
\psfrag{w7}[c][c][0.9]{$w_{7}$}
\psfrag{w8}[c][c][0.9]{$w_{8}$}
\psfrag{w9}[c][c][0.9]{$w_{9}$}
\psfrag{w10}[c][c][0.9]{$w_{10}$}
\psfrag{w11}[c][c][0.9]{$w_{11}$}
\psfrag{w12}[c][c][0.9]{$w_{12}$}
\psfrag{w13}[c][c][0.9]{$w_{13}$}
\psfrag{w14}[c][c][0.9]{$w_{14}$}
\psfrag{w15}[c][c][0.9]{$w_{15}$}
\psfrag{w16}[c][c][0.9]{$w_{16}$}
\psfrag{w17}[c][c][0.9]{$w_{17}$}
\psfrag{w18}[c][c][0.9]{$w_{18}$}
\psfrag{w19}[c][c][0.9]{$w_{19}$}
\psfrag{w20}[c][c][0.9]{$w_{20}$}
\psfrag{w21}[c][c][0.9]{$w_{21}$}
\includegraphics[scale=0.60, angle=-90]{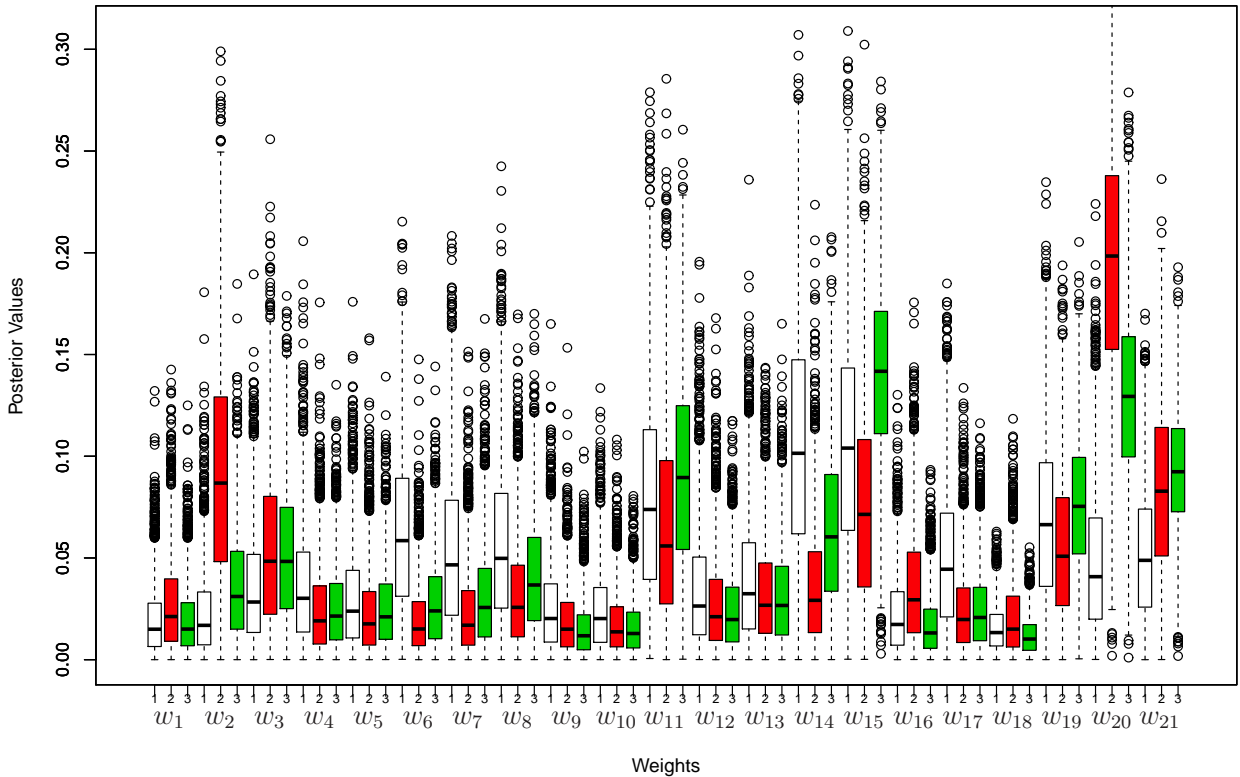}
\end{center}
\vspace{-2em}
\end{figure}

\begin{figure}[ht!]
\caption{Boxplots summarizing posterior distributions of the differences in the model weights (First Period - Second Period)}
\label{comp2}
\vspace{-2em}
\begin{center}
\psfrag{w1}[c][c][0.9]{$w_{1}$}
\psfrag{w2}[c][c][0.9]{$w_{2}$}
\psfrag{w3}[c][c][0.9]{$w_{3}$}
\psfrag{w4}[c][c][0.9]{$w_{4}$}
\psfrag{w5}[c][c][0.9]{$w_{5}$}
\psfrag{w6}[c][c][0.9]{$w_{6}$}
\psfrag{w7}[c][c][0.9]{$w_{7}$}
\psfrag{w8}[c][c][0.9]{$w_{8}$}
\psfrag{w9}[c][c][0.9]{$w_{9}$}
\psfrag{w10}[c][c][0.9]{$w_{10}$}
\psfrag{w11}[c][c][0.9]{$w_{11}$}
\psfrag{w12}[c][c][0.9]{$w_{12}$}
\psfrag{w13}[c][c][0.9]{$w_{13}$}
\psfrag{w14}[c][c][0.9]{$w_{14}$}
\psfrag{w15}[c][c][0.9]{$w_{15}$}
\psfrag{w16}[c][c][0.9]{$w_{16}$}
\psfrag{w17}[c][c][0.9]{$w_{17}$}
\psfrag{w18}[c][c][0.9]{$w_{18}$}
\psfrag{w19}[c][c][0.9]{$w_{19}$}
\psfrag{w20}[c][c][0.9]{$w_{20}$}
\psfrag{w21}[c][c][0.9]{$w_{21}$}
\includegraphics[scale=0.60, angle=-90]{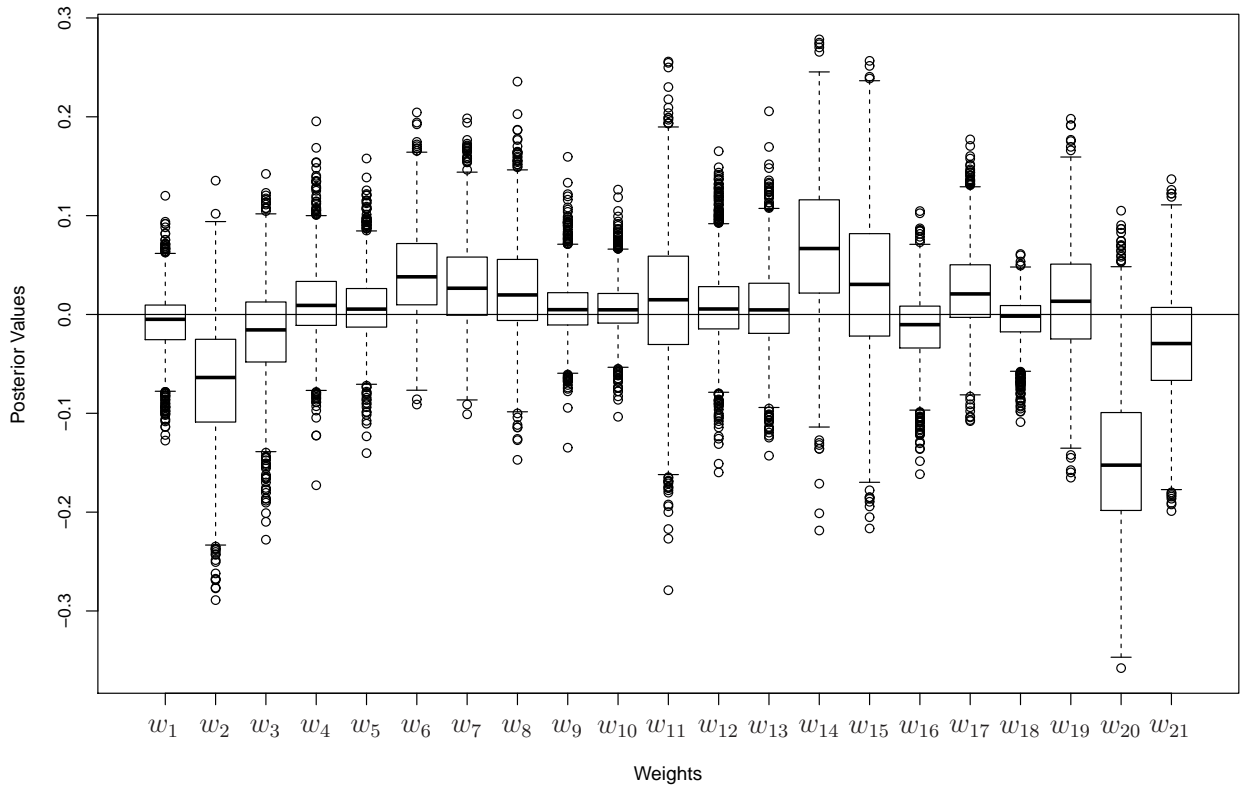}
\end{center}
\vspace{-2em}
\end{figure}

Initially we fit (for $j=1,2$) model (\ref{sep}). We run the MCMC algorithm for 3000 iterations, with the first 1000 used as the burnin period. The DIC for this model was  -329.614. From Figure \ref{comp} we observe some differences between the posterior distributions of some weights. In particular weights 2, 6, 14 and 20 seem to differ between the two time periods. We draw the same conclusions from Figure \ref{comp2}, where we have presented boxplots that summarize posterior distributions of the differences (period 1 - period 2) in the model weights. 

Table \ref{sm} presents posterior means for the weights on the two different time periods as well as posterior means for the differences in weights (first period - second period). Additionally in the fourth column of Table \ref{sm} we have placed the tail-area probabilities of zero for the weight differences, see for example \citet[page 155]{ntzoufras_2009}, i.e. the posterior probability  
$$
\pi_0=\min\left\{f(w_{1,\ell}-w_{2,\ell}>0|\mathbf{y}),~f(w_{1,\ell}-w_{2,\ell}<0|\mathbf{y}) \right\},~~\ell=1,\dots,21.
$$
When the zero value lies at the center of the posterior distribution, then the value of $\pi_0$ is expected to be close to 0.5, indicating that there are no clear differences of the weights between the two different time periods. When $\pi_0$ is low (e.g. lower than 5\%) then we can conclude that there are differences of the weights between the two different time periods. Again for weights 2, 6, 14 and 20 the value of $\pi_0$ is below 20\% and especially for weight 20 is around 3\%. In the Discussion Session we explain those observed differences.    

Using the same number of iterations and burnin period we also run the MCMC algorithm for model (\ref{joint}). The DIC for this model was -331.751, indicating a slightly worse fit than before. However, we have to mention here that the posterior distributions of the weights are not completely symmetric, and therefore DIC must be used with caution. We have also considered a simple $R^2$ type measure that measures the proportion of explained variation relative to total variation. For both models the posterior mean of this statistic is around 0.65, which suggests that there is no need to use the more complicated model (different time periods) over the simpler (joint model). 

Table \ref{jm} presents posterior summaries for the weights when using the joint model. Weights 3, 11, 14, 15, 19, 20 and 21 have posterior means above 5\%. From the results in Table \ref{jm} we conclude that students give the higher relative importance to attribute number 15 (with posterior mean weight around 14\%), followed by attribute number 20 (with posterior mean weight around 13\%) and attribute number 11 (with posterior mean around 9\%). Also the weight of the latent variable has posterior mean around 9\%, i.e. the relative importance that students give to unobserved random variables in determining the overall evaluation score is around 9\%. 

Attribute number 15 refers to whether lecturers have adequate and in depth knowledge of their subjects.  Attribute 20 refers to whether lecturers try to provide through their courses extension of their subjects to current scientific and actual trends. Attribute 11 refers to whether lecturers organize well their courses (content, teaching, exams). We can therefore conclude that students constantly pay high importance to the core issues related to the quality of their academic studies: academic quality of their lecturers, quality of their courses and degree of connection of their courses to contemporary scientific developments.

The relative importance of attributes 15, 11 and of the latent variable is almost the same between the two studied periods. On the other hand for attribute 20 we observe major differences. While for the first period studied this attribute had a weight around 5\%, for the second period the corresponding weight has increased noticeably and has posterior mean above 19\%.

\begin{table}[ht!]
\caption{Posterior means for the weights of the two separated models and for the weight differences (first period - second period). The last column presents tail-area probabilities of zero for the weight differences}
\label{sm}
\begin{center}
\begin{tabular}{ccccc}
\hline Weight & First Period & Second Period & Differences & $\pi_0$ \\
\hline                                                 
  $w_1$&	0.0198&	0.0278	&	-0.0079&	0.4050\\	
	$w_2$&	0.0233&	0.0926	&	-0.0693&	0.1155\\
	$w_3$&	0.0357&	0.0552	&	-0.0194&	0.3675\\
	$w_4$&	0.0369&	0.0254	&	0.0114&	0.3860\\
	$w_5$&	0.0305&	0.0233	&	0.0071&	0.4150\\
	$w_6$&	0.0626&	0.0203	&	0.0423&	0.1645\\
	$w_7$&	0.0535&	0.0237	&	0.0298&	0.2580\\
	$w_8$&	0.0572&	0.0326	&	0.0246&	0.3095\\
	$w_9$&	0.0260&	0.0194	&	0.0066&	0.4220\\
	$w_{10}$&	0.0247&	0.0184&	0.0062&	0.4205\\
	$w_{11}$&	0.0808&	0.0668	&	0.0139&	0.4065\\
	$w_{12}$&	0.0355&	0.0278	&	0.0077&	0.4220\\
	$w_{13}$&	0.0401&	0.0335	&	0.0065&	0.4460\\
	$w_{14}$&	0.1062&	0.0371	&	0.0691&	0.1445\\
	$w_{15}$&	0.1067&	0.0765	&	0.0301&	0.3425\\
	$w_{16}$&	0.0227&	0.0362	&	-0.0135&	0.3595\\	
	$w_{17}$&	0.0500&	0.0253	&	0.0246&	0.2760\\
	$w_{18}$&	0.0155&	0.0213	&	-0.0057&	0.4610\\
	$w_{19}$&	0.0698&	0.0561	&	0.0137&	0.4075\\
	$w_{20}$&	0.0486&	0.1957	&	-0.1471&	0.0305\\
	$w_{21}$&	0.0527&	0.0838	&	-0.0310&	0.2970\\               
 \hline
\end{tabular}
\end{center}
\end{table}

\begin{table}[ht!]
\caption{Posterior summaries for the weights of the joint model}
\label{jm}
\begin{center}
\begin{tabular}{llllll}
\hline
 &  &   & 2.5\%  &   & 97.5\%   \\
Weight&Mean &SD & Percentile &Median &Percentile \\
\hline
  $w_1$&	0.0199&	0.0176	&	0.0007&	0.0150&	0.0659\\	
	$w_2$&	0.0368&	0.0277	&	0.0014&	0.0310&	0.1039	\\
	$w_3$&	0.0525&	0.0342	&	0.0032&	0.0483&	0.1291	\\
	$w_4$&	0.0263&	0.0211	&	0.0009&	0.0214&	0.0799	\\
	$w_5$&	0.0258&	0.0205	&	0.0009&	0.0210&	0.0760	\\
	$w_6$&	0.0282&	0.0220	&	0.0010&	0.0239&	0.0819	\\
	$w_7$&	0.0311&	0.0253	&	0.0009&	0.0256&	0.0956	\\
	$w_8$&	0.0417&	0.0287	&	0.0023&	0.0367&	0.1068	\\
	$w_9$&	0.0154&	0.0139	&	0.0003&	0.0117&	0.0523	\\
	$w_{10}$&	0.0163&	0.0135&	0.0004&	0.0128&	0.0489	\\
	$w_{11}$&	0.0911&	0.0481	&	0.0079&	0.0895&	0.1905	\\
	$w_{12}$&	0.0248&	0.0205	&	0.0006&	0.0197&	0.0767	\\
	$w_{13}$&	0.0318&	0.0250	&	0.0009&	0.0266&	0.0899	\\
	$w_{14}$&	0.0650&	0.0396	&	0.0045&	0.0603&	0.1526	\\
	$w_{15}$&	0.1402&	0.0451	&	0.0501&	0.1418&	0.2257\\
	$w_{16}$&	0.0172&	0.0151	&	0.0005&	0.0131&	0.0566\\	
	$w_{17}$&	0.0251&	0.0203	&	0.0008&	0.0208&	0.0782	\\
	$w_{18}$&	0.0120&	0.0093	&	0.0004&	0.0101&	0.0346	\\
	$w_{19}$&	0.0758&	0.0344	&	0.0094&	0.0753&	0.1441	\\
	$w_{20}$&	0.1296&	0.0439	&	0.0423&	0.1294&	0.2131	\\
	$w_{21}$&	0.0925&	0.0309	&	0.0303&	0.0924&	0.1520	\\
\hline
\end{tabular}
\end{center}
\end{table}

\section{Discussion}

Since the two data sets have a good fit, it implies that the model provides a satisfactory reflection of the way that university students evaluate their program of studies. However, the boxplots of the model weights presented in Figure \ref{comp} lead us to the following comments. Between 2009 and 2013 we observe a number of differences in the weights of our model. Some of them could be considered as notable while some others as minor:

\begin{itemize}

\item[A.] We observe the existence of notable differences in weights $w_2$, $w_6$, $w_{14}$ and $w_{20}$. More specifically:
\begin{enumerate}
\item	The weight attached to question or variable 2 ($w_2$) had increased in 2013 compared to 2009. That is students paid more importance on the issue whether the syllabus of their courses had been in accordance to their objectives. 
\item	The weight attached to question 6 ($w_6$) had been reduced in 2013 compared to 2009. That is students considered course attendance as a factor of diminishing importance for the understanding and the successful completion of their courses.
\item	The weight attached to question 14 ($w_{14}$) had been reduced in 2013 compared to 2009. That is students considered the communicability of their lecturers as a diminishing importance factor affecting their decisions to lecture participating.
\item	The weight attached to question 20 ($w_{20}$) had increased in 2013 compared to 2009. That is students paid more importance on the issue whether their lecturers tried to provide through their courses extension of their subjects to current scientific and actual trends. 
\end{enumerate}
\item[B.] We observe the existence of minor differences in weights $w_4$, $w_7$, $w_8$, $w_{15}$, $w_{17}$ and $w_{21}$. More specifically: The weights attached to questions 4, 7, 8, 15 and 17, that is the values of $w_4$, $w_7$, $w_8$, $w_{15}$ and $w_{17}$, had been reduced between 2009 and 2013, while the weight attached to question 21, that is the value of $w_{21}$, had increased. Although all these changes were minor, taken together they lead us to the interesting conclusion that the importance of those factors that shape the general academic environment of studies (the quality of teaching material, the way of exams, the compatibility between the course and its exams, the level of the lecturer's knowledge of the subject, the consistency of the lecturer with his obligations referring to his students) has been reduced during the above period, although at the same time students consider that the quality of teaching in the department has been improved ($y$). It is not therefore the case that the diminishing importance of the academic environment could be attributed to a possible deterioration of the quality of the department's academic staff, but to other factors.
\end{itemize}

Before we discuss further the outcomes of our statistical analysis, we consider it necessary to provide some insight to the considerable changes during the period 2009-2013 of some factors or indicators that shape the general socio-economic environment in Greece. According to Eurostat estimates, during the period 2009-2013 real GDP per capita in Greece had been reduced by 26.2 percentage units, that is on average Greeks lost more than $1/4$ of their real incomes. It has been estimated that real GDP per capita in Greece in 2013 had been reduced to its pre 2000 levels. During the period 2009-2012 (since no data or estimates exist yet for 2013) the total number of unemployed members of the labour force increased from 471 to 1204 thousand people, that is more than 150\% (155.6\%) or from 9.5\% to 24.3\% of the labour force. The unemployment rate of labour force aged less than 25 years had been increased from 25.8\% to 55.3\%. This young unemployment rate was the highest in the European Union.

During the same period 2009-2012 the number of people at risk of poverty or social exclusion, that is the sum of persons who are at risk of poverty or severely materially deprived or living in households with very low work intensity, is estimated to had increased in Greece from 3007 to 3795 thousand people or from 27.6\% to 34.6\% of total population. More specifically, the number of severely deprived people had increased from 1198 to 2141 thousand, that is almost by 80\% (78.7\%), while the number of people living in households with very low work intensity (people aged 0-59 living in households where the adults work less than 20\% of their total work potential during the past year) increased from 539 to 1158 thousand, that is almost by 115\% (114.8\%). It is important to note that during the same period, the percentage of people at risk of poverty by highest level of education attained (first and second stage of tertiary education, i.e. levels 5 and 6) increased from 5.3\% to 9.4\%. The risk of poverty rate of 9.4\% of this particular educational group was the highest in the European Union. 

The time path of all the above socio-economic indicators in Greece during the period 2004-2013 as they are provided by the Eurostat (where data exist otherwise 2004-2012) is presented in Table \ref{disc_tab}. It is expected that during the year 2013 all the above socio-economic factors had been aggravated, since real GDP per capita is estimated to had been reduced further by 4.2 percentage units and in July 2013 total unemployment had further increased to 1374 thousand people or to 27.6\% of the labour force. 

\begin{table}[ht!]
\scriptsize
\caption{The development of some characteristic socio-economic indicators in Greece during the period 2004-2013}
\label{disc_tab}
\begin{center}
\tabcolsep=0.11cm
\begin{tabular}{ccccccccccc}
\hline
Year &	Real & Real & Unemploy- & Total & Unemploy- & People & People& Severely & People& At-risk-\\ 
& GDP & GDP & ment & unemploy- & ment & at risk of& at risk of & materially & living in & of-poverty-\\
& per & per & (in 1000 & ment& rate of& poverty& poverty& deprived& households & rate\\
& capita& capita& persons) &  rate & labour & or social & or social & people & with very & (first and \\
& (in \euro & (\% change & & (\% of  & force aged & exclusion & exclusion & (in 1000 & low work & second\\ 
& per & on  & & labour & less than & (in 1000 & (\% of total & persons) & 
 intensity & stage of \\
& inhabi-& previous&& force)&25 years & persons)& population)&& (in 1000 & tertiary\\
& tant) & year) & & & (\% of & & & & persons)&education,\\
&       &       & & &  correspo-& & & & & i.e. levels\\
&       &       & & &  nding    & & & & & 5 and 6)\\
&       &       & & &  labour   & & & & & \\
&       &       & & &  force) & & & & & \\
\hline
2004&	17100&	4.0&	506&	10.5&	26.9&	3283&	30.9&	1500&	604&	5.1\\
2005&	17400&	1.9&	477&	9.9&	26.0&	3131&	29.4&	1365&	610&	5.6\\
2006&	18300&	5.1&	434&	8.9&	25.2&	3154&	29.3&	1236&	660&	5.7\\
2007&	18800&	3.1&	407&	8.3&	22.9&	3064&	28.3&	1238&	662&	7.5\\
2008&	18700&	-0.6&	378&	7.7&	22.1&	3046&	28.1&	1213&	611&	6.8\\
2009&	18100&	-3.5&	471&	9.5&	25.8&	3007&	27.6&	1198&	539&	5.3\\
2010&	17100&	-5.2&	629&	12.6&	32.9&	3031&	27.7&	1269&	619&	5.8\\
2011&	15900&	-7.0&	877&	17.7&	44.4&	3403&	31.0&	1667&	978&	7.1\\
2012&	14900&	-6.3&	1204&	24.3&	55.3&	3795&	34.6&	2141&	1158&	9.4\\
2013&	     &	-4.2&	    &	    &	    &     &	    &	    &	    &	   \\
\hline
\end{tabular}
\end{center}
\footnotesize
\textit{Source:} Eurostat: \url{http://epp.eurostat.ec.europa.eu} and \url{http://appsso.eurostat.ec.europa.eu}. Date of extraction: 22 Oct. 2013.
\end{table}

\normalsize

The development of all the above indicators clearly depicts the great socio-economic transformation that Greece was experiencing during the period 2009-2013, mainly as a consequence of the policies adopted to handle its public debt crisis that caused an unprecedented economic crisis. These crises have affected both the supply and the demand side of education. The supply side has been affected as a result of the large cuts on public expenditure on education, while the demand side has been affected mainly by the considerable reductions on individual disposable incomes and the high levels of unemployment, even among the university graduates. Young people realized suddenly that university degrees cannot provide a stable security against unemployment as it was widely believed in the past and at the same time many of them do not have the necessary means to support their studies. 

It is not therefore surprising that this transformation has affected the criteria by which students evaluate their university studies and consequently the targets they pursue during their study years. They have shifted their interests from the academic environment of their studies (course attendance, communicability of lecturer, quality of teaching material, way of exams, compatibility between courses and their exams, level of lecturer knowledge, consistency of lecturer) to the more up to date developments of their courses (courses related to modern scientific developments/timeliness) and the consistency between syllabus and its targets, possibly believing that in that way they are going to acquire a comparative advantage to find a job relatively to past graduates with rather obsolete knowledge.

We distinguish therefore two trends: On the one side we have the deterioration of the importance of the academic environment or academic life and on the other the increasing importance of the consistency of studies and their modern developments. The deterioration of academic environment could be attributed to the fact that possibly many students lack the opportunity to attend their courses, either because they have to work to support themselves and their families or they cannot afford the expenses of academic life (living, transport and other costs), especially if they live in another area. 

Obviously, this change of priorities is going to have large long-run implications on the characteristics of the Greek system of higher education and on the process of the Greek socio-economic development, since many students who cannot afford to attend their courses will turn to a form of unofficial distance learning, facing no equal study opportunities.

We must point out that we were not able to find similar research to compare the results of our analysis since all the existing empirical work is limited to the effects of economic crises on public education expenditure.

\section{Conclusion}

To conclude with, since every evaluation is or should be a continuous, consistent and user friendly process, in this paper we have proposed a Bayesian hierarchical model which provides:
\begin{itemize}
\item user friendly inferences;
\item direct probability interpretations;
\item natural ways for implementation of restrictions; 
\item capability for prior information incorporation;
\item capability for continuous assessment and monitoring.
\end{itemize}
These characteristics indicate possible extensions/applications of the proposed model to longitudinal studies that besides the continuous assessment and monitoring of service quality, will examine the association of customers preferences with various socio economic time series.


\bibliographystyle{agsm}

\bibliography{biblio3}

\end{document}